\documentstyle[11pt, paspconf]{article}

\begin{document}

\title{A free-floating planet population in the Galaxy?}

\author{Hans Zinnecker \altaffilmark{}} 

\affil{Astrophysikalisches Institut Potsdam, An der Sternwarte 16,\\
       D-14482 Potsdam, Germany (hzinnecker@aip.de)}

\begin{abstract}
Most young low-mass stars are born as binary systems, and circumstellar
disks have recently been observed around the individual components of
proto-binary systems (e.g. L1551-IRS5). Thus planets and planetary systems
are likely to form around the individual stellar components in sufficiently
wide binary systems. However, a good fraction of planets born in binary
systems will in the long run be subject to ejection due to gravitational
perturbations. Therefore, we expect that there should exist a free-floating
population of Jupiter-like or even Earth-like planets in interstellar space.
There is hope to detect the free-floating Jupiters through gravitational
microlensing observations towards the Galactic Bulge, especially with 
large-format detectors
in the near-infrared (e.g. with VISTA or NGST), 
on timescales of a few days.\footnote{Contributed talk, 
 presented at ``Microlensing 2000", Cape Town (Feb. 2000), in press} 

\end{abstract}

% Keywords should be included, but they are not printed in the hardcopy.


\section{Speculation of Existence}
\par
Observations over the past decade have shown that most young low-mass
stars are born as binary systems (e.g. Mathieu 1994, Mayor et al. 2000).
Furthermore, circumstellar disks have recently been observed around the
individual stellar components of protobinary systems (e.g. L1551-IRS5,
Rodriguez et al. 1998; HK Tau, Menard \& Stapelfeldt 2000). Thus, planets
and planetary systems are likely to form around the individual components
in sufficiently wide binary systems. A case in point is 16 Cygni, where
a Jupiter-like companion has been detected around component B by radial
velocity measurements (Cochran et al. 1996).

Here we speculate that a good fraction of planets born in binary systems
will be subject to dynamical ejection due to gravitational perturbations
resulting from periastron passages of the stars revolving around each other,
typically in rather eccentric orbits.
\footnote {Although the case of TMR-1C (Terebey et al. 2000) did not 
turn out to be a young planet being ejected, some
young multiple systems appear 
to eject very low-mass objects (Reipurth 2000).}
Therefore we expect a population of
ejected and thus free-floating planets to exist in the Galactic Disk -- and
maybe in the Galactic Halo as well, as the frequency of wide visual binaries
(separations ~$>$~ 30 AU) seems to be at least as high for halo stars
as for disk stars
(K\"ohler et al. 2000).

\section{Caveat}
To be fair, there are also arguments against our speculation. For example,
planet formation may be inhibited by the mutual spiral shocks induced in
the respective circumstellar disks which will heat the disk gas and dust
(Nelson 2000). This might cause the dust grains to evaporate and be 
destroyed, certainly the ice mantles if not the silicate cores too,
depending on the shock temperature (higher/lower than 1500K).Thus grain
growth and the collisional build-up of planetesimals will be prevented.
Similarly, direct gravitational instability of the gas disk (another
possibility to form a giant Jupiter-like planet) is impeded, if the disk
is too hot (the Toomre Q parameter exceeds unity for too large a sound
speed or gas temperature, even if the disk surface density is high).
Finally, the Goldreich-Ward (1973) instability of a cold thin dust disk,
even if it operated in a circumstellar disk around a single star, may not
do so in the respective disks around binary star components, as the dust
may never settle into the disk's midplane in a binary star+disk system.

While these qualitative
arguments against planet formation in binary systems must be
further investigated and while a lot of uncertainty remains, it is 
nonetheless
worthwhile to proceed on the assumption that some planet formation
occurs even in binary systems, especially in those which are wider than
a critical separation of the stellar pair, which we take to be the 
separation where the mode of the semi-major axis distribution
occurs -- 30 AU according to Duquennoy \& Mayor (1991). That is, we assume
that 50\% of all low-mass binaries can indeed form planets
(cf. Marzari \& Scholl 2000).

\section {Prospects for Detection}

Next we evaluate the prospects for detecting a population of Jupiter-mass
free-floating planets towards the Galactic Bulge (Center). We also bracket our 
estimates by considering objects a factor of 10 more massive (i.e.
minimum mass brown dwarfs) and a factor 10 less massive (maximum mass of
Earth-type planets).
We consider planets associated with stars below 1 $M_{\odot}$
(with main sequence lifetimes exceeding the age of the Galaxy). We assume
50\% binary frequency for these low-mass stars, half of which we expect 
to form planets. Each planet forming binary system is assumed to form
4 planets, two around each component, on average. Finally we assume
that one of the two planets per stellar component is eventually ejected.

The number density of low-mass stellar systems in the solar neighborhood
is $n_{*}$ = 0.1 per pc$^{3}$. Thus, under the above assumption, the 
number density of free-floating planets (Jupiters) will be $n_{p}$ = 0.05
per pc$^{3}$. Then the surface density of free-floating planets towards
the Galactic Center will be of order $N_{p}$ = 10$^{3}$ pc$^{-2}$,
assuming a distance to the Galactic Center of 10 kpc
and an average stellar density in the inner Galaxy
2 times higher than in the local Solar Neighborhood. 

Now, the probability P (also sometimes called the ``optical depth") for
a microlensing occurrence is given by the following area coverage factor

\vskip0.3cm

P = $\pi ~\times$~$R^{2}_{E,p}$ $\times$~ N$_{p}$

\noindent
where $R_{E,p}$ is the Einstein radius of the planet  
($R_{E,p}$ = $\sqrt{4GM_{p}D_{s}x(1-x)/c^{2}}$, with $D_{s}$
being the distance to the source population, here of order 10 kpc;
and x=$D_{L}/D_{s}$, the ratio of the distance of the lens
population to the source population, typically x=0.5).

When normalized to $M_{Jup}$ = 10$^{-3}$
M$_{\odot}$ and for x=0.5, numerically this turns out to be\\

$R_{E,p}$ = 10$^{12.5}$ $\times$ $\sqrt{M/M_{Jup}}$   cm.\\
 
Hence\\ 

P = 3 $\times$ 10$^{-9}$ $\times$~(M/$M_{Jup}$)
\vskip0.3cm
\noindent
is a reasonable estimate for the ``optical depth" due to
free-floating planets of mass M towards the inner Galaxy.

\noindent
The timescale for Einstein ring crossing, i.e. the half-width
of the lightcurve of the source magnification, is given by\\ 

t = $R_{E,p}$/v

\vskip0.3cm
\noindent
where v $\sim$ 200 km/s is the relative speed of the observer
with respect to the lens.

One finds\\

t = 10$^{12.5}$/10$^{7.3}$ sec = 1.5 $\times$ 10$^{5}$ sec~ $\sim~$ 2 days

\section {Detection Requirements}

\par
The above numbers translate into the following detection requirements:
to detect a microlensing event due to free-floating Jupiters in the
inner Galaxy we need to observe some 3 $\times$ 10$^{8}$ stellar objects
in the Galactic Bulge for a few days, with a time resolution of hours.
This is difficult but not impossible.
For example, the VST 2.5m telescope to be installed on Paranal/Chile
in 2002 with its 1 degree field-of-view in very good seeing (0.4 arcsec)
corresponds to $\sim$10$^{8}$ pixels, and may thus be able to
realistically resolve some 10$^{7}$ point sources down to the confusion 
limit in the far-red bands (10 pixels per object).
Therefore some 30 fields of 1 degree
must be monitored each night.
The VISTA telescope, a 4m mirror on Paranal with a 1 degree field-of-view
for infrared (JHK) observations, will be even better (first light in 2005)
because in the near-infrared we can penetrate the dust towards the 
Galactic Center, providing a higher and fainter source surface density.
Thus the confusion limit will be reached more quickly and 30 fields
can actually be monitored once or twice per night (good sampling).
Finally, it is conceivable that the Next Generation Space Telescope (NGST),
to be launched in 2009, will be able to do the job, as it will mostly operate
in staring mode and always at the diffraction limit (60 mas). Its 
field-of-view is around 4 arcmin (8k x 8k detectors with 30 mas pixels),
corresponding to 10$^{7.5}$ pixels. Due to the much higher spatial
resolution compared to VISTA, the NGST will reach a fainter confusion
limit than VISTA within a short exposure time (a few seconds, or tens
of seconds, depending on the precise galactic longitude and latitude)
and will likely resolve 10$^{6.5}$ point sources.
It then 'just' needs 100 such short exposures of the same field, 
sampled at intervals of a few hours,
in order to monitor three hundred million objects for 
a microlensing variability of a few days, thus 
finding out if the predicted microlensing by free-floating Jupiters occurs.
If so, this would herald a new class of objects in interstellar space.

PS. Microlensing constraints on the frequency of Jupiter-mass planets
{\it in orbit} around low-mass stars with separations of
1.5 to 3 AU (the lensing zone) were analyzed by the PLANET collaboration.
No clear signatures of such planets among 100 microlensing events
have been detected, implying a frequency of less than 33\% 
(Gaudi et al. 2000, these Proceedings). Bound Jupiters in wider orbits
(wider than 1.5 Einstein radii or about 5 AU for a solar-mass lens)
are much harder to distinguish from free-floating Jupiters, with only 
a small fraction of microlensing events still carrying a weak signal 
of the parent star (cf. DiStefano \& Scalzo 1999; Gaudi, priv. comm.).

\section {Acknowledgement}
My first attempt to estimate the number of free-floating planets
ejected from binary systems and their detection by gravitational 
microlensing goes back to a young binary stars workshop in Stony Brook 1996.
Since then, I benefitted enormously from the expertise of my colleague
Joachim Wambsganss to whom I owe both my increasing
interest and increasing knowledge of gravitational microlensing.
I am also grateful to Rosanne DiStefano for insightful discussions.


\begin{references}

\reference
Cochran, Hatzes, Marcy \& Butler 1996, BAAS 28, 1111 

\reference
DiStefano \& Scalzo 1999, ApJ. 512, 564 

\reference
Duquennoy \& Mayor 1991, A \& A 248, 485

\reference
Gaudi et al. 2000, astro-ph/0004269

\reference
Goldreich \& Ward 1973, ApJ. 183, 1051 

\reference
K\"ohler, Zinnecker, \& Jahreiss 2000, in 
Birth and Evolution of Binary Stars (Poster Proc. IAU-Symp. 200),
eds. Reipurth \& Zinnecker, p. 148

\reference
Marzari \& Scholl 2000, ApJ. 543, 328

\reference
Mathieu 1994, ARAA 32, 465

\reference
Mayor et al. 2000, in 
The Formation of Binary Stars, IAU Symp. 200, \\    
eds. Zinnecker \& Mathieu, ASP Conf. Series, in press

\reference
Menard \& Stapelfeldt 2000, in
The Formation of Binary Stars, IAU Symp. 200, \\     
eds. Zinnecker \& Mathieu, ASP Conf. Series, in press

\reference
Nelson 2000, ApJ. 537, L65 

\reference
Reipurth 2000, in
The formation of Binary Stars, IAU Symp. 200,\\
eds. Zinnecker \& Mathieu, ASP Conf. Series, in press

\reference
Rodriguez et al. 1998, Nature 395, 355

\reference
Terebey et al. 2000, A.J. 119, 2341 

\end{references}
\end{document}